\documentclass[12pt]{article}
\usepackage{graphicx}

\begin{document}

\title{\textbf{Production of heavy and superheavy nuclei in massive fusion reactions}}

\author{Zhao-Qing Feng$^{1}$\footnote{Corresponding author. \newline \emph{E-mail address:} fengzhq@impcas.ac.cn},
Gen-Ming Jin$^{1}$, Jun-Qing Li$^{1}$, Werner Scheid$^{2}$}
\date{}
\maketitle

\begin{center}
$^{1}${\small \emph{Institute of Modern Physics, Chinese Academy of
Sciences, Lanzhou 730000, China}}\\[0pt]
$^{2}${\small \emph{Institut f\"{u}r Theoretische Physik der
Universit\"{a}t, 35392 Giessen, Germany}}
\end{center}

\textbf{Abstract}
\par
Within the framework of a dinuclear system (DNS) model, the
evaporation-residue excitation functions and the quasi-fission mass
yields in the $^{48}$Ca induced fusion reactions are investigated
systematically and compared with available experimental data.
Maximal production cross sections of superheavy nuclei based on
stable actinide targets are obtained. Isotopic trends in the
production of the superheavy elements Z=110, 112-118 based on the
actinide isotopic targets are analyzed systematically. Optimal
evaporation channels and combinations as well as the corresponding
excitation energies are proposed. The possible factors that
influencing the isotopic dependence of the production cross sections
are analyzed. The formation of the superheavy nuclei based on the
isotopes U with different projectiles are also investigated and
calculated.

\section{Introduction}

The synthesis of heavy or superheavy nuclei is a very important
subject in nuclear physics motivated with respect to the island of
stability which is predicted theoretically, and has obtained much
experimental research with the fusion-evaporation reactions
$\cite{Ho00,Og07}$. The existence of the superheavy nucleus (SHN)
($Z\geq106$) is due to strong binding shell effects against the
large Coulomb repulsion. However, the shell effects get reduced with
increasing the excitation energy of the formed compound nucleus.
Combinations with a doubly magic nucleus or nearly magic nucleus are
usually chosen owing to the larger reaction $Q$ values. Reactions
with $^{208}$Pb or $^{209}$Bi targets were first proposed by
Oganessian et al. to synthesize SHN $\cite{Og75}$. Six new elements
with Z=107-112 were synthesized in cold fusion reactions for the
first time and investigated at GSI (Darmstadt, Germany) with the
heavy-ion accelerator UNILAC and the SHIP separator
$\cite{Ho00,Mu99}$. Recently, experiments on the synthesis of
element 113 in the $^{70}$Zn+$^{209}$Bi reaction have been performed
successfully at RIKEN (Tokyo, Japan) $\cite{Mo04}$. However, it is
difficulty to produce heavier SHN in the cold fusion reactions
because of the smaller production cross sections that are lower than
1 pb for $Z>113$. Other possible ways to produce SHN are very needed
to be investigated in experimentally and theoretically. Recently,
the superheavy elements Z=113-116, 118 were synthesized at FLNR in
Dubna (Russia) with the double magic nucleus $^{48}$Ca bombarding
actinide nuclei $\cite{Og99,Og04a,Og06}$. New heavy isotopes
$^{259}$Db and $^{265}$Bh have also been synthesized at HIRFL in
Lanzhou (China) $\cite{Ga01}$. Further experimental works are
necessary in order to testify the new synthesized SHN. A reasonable
understanding of the formation of SHN in the massive fusion
reactions is still a challenge for theory.

In accordance with the evolution of two heavy colliding nuclei, the
dynamical process of the compound nucleus formation and decay is
usually divided into three reaction stages, namely the capture
process of the colliding system to overcome the Coulomb barrier, the
formation of the compound nucleus to pass over the inner fusion
barrier, and the de-excitation of the excited compound nucleus by
neutron emission against fission. The transmission in the capture
process depends on the incident energy and relative angular momentum
of the colliding nuclei, which is the same as that in the fusion of
light and medium mass systems. The complete fusion of the heavy
system after capture in competition with quasi-fission is very
important in the estimation of the SHN production. The concept of
the "extra-push" energy explains for the fusion of two heavy
colliding nuclei in the macroscopic dynamical model
$\cite{Sw80,Bj82}$. At present it is still difficult to make an
accurate description of the fusion dynamics. After the capture and
the subsequent evolution to form the compound nucleus, the thermal
compound nucleus will decay by the emission of light particles and
$\gamma$ rays against fission. The three stages will affect the
formation of evaporation residues observed in laboratories. The
evolution of the whole process of massive heavy-ion collisions is
very complicated at near-barrier energies. Most of the theoretical
methods on the formation of SHN have a similar viewpoint in the
description of the capture and the de-excitation stages, but there
are different description of the compound nucleus formation process.
There are mainly two sorts of models, depending on whether the
compound nucleus is formed along the radial variable (internuclear
distance) or by nucleon transfer in a touching configuration which
is usually the minimum position of the interaction potential after
capture of the colliding system. Several transport models have been
established to understand the fusion mechanism of two heavy
colliding nuclei leading to SHN formation, such as the macroscopic
dynamical model $\cite{Sw80,Bj82}$, the fluctuation-dissipation
model $\cite{Ar99}$, the concept of nucleon collectivization
$\cite{Za01}$ and the dinuclear system model $\cite{Ad97,Fe06}$.
Recently, the improved isospin-dependent quantum molecular dynamics
(ImIQMD) model was also proposed to investigate the fusion dynamics
of SHN $\cite{Wa04,Fe08}$. With these models experimental data can
be reproduced to a certain extent, and some new results have been
predicted. However, these models differ from each other, and
sometimes different physical ideas are used.

Further improvements of these models have to be made. Here we use a
dinuclear system (DNS) model $\cite{Fe06,Fe07}$, in which the
nucleon transfer is coupled with the relative motion by solving a
set of microscopically derived master equations, and a barrier
distribution of the colliding system is introduced in the model. We
present a new and extended investigation of the production of
superheavy nuclei in the $^{48}$Ca induced fusion reactions and in
other combinations.

In Section 2 we give a simple description on the DNS model.
Calculated results of fusion dynamics and SHN production are given
in Section 3. In Section 4 conclusions are discussed.

\section{Dinuclear system model}

The dinuclear system $\cite{Vo78}$ is a molecular configuration of
two touching nuclei which keep their own individuality
$\cite{Ad97}$. Such a system has an evolution along two main degrees
of freedom: (i) the relative motion of the nuclei in the interaction
potential to form the DNS and the decay of the DNS (quasi-fission
process) along the R degree of freedom (internuclear motion), (ii)
the transfer of nucleons in the mass asymmetry coordinate
$\eta=(A_{1}-A_{2})/(A_{1}+A_{2})$ between two nuclei, which is a
diffusion process of the excited systems leading to the compound
nucleus formation. Off-diagonal diffusion in the surface $(A_{1},R)$
is not considered since we assume the DNS is formed at the minimum
position of the interaction potential of two colliding nuclei. In
this concept, the evaporation residue cross section is expressed as
a sum over partial waves with angular momentum $J$ at the
centre-of-mass energy $E_{c.m.}$,
\begin{equation}
\sigma_{ER}(E_{c.m.})=\frac{\pi \hbar^{2}}{2\mu
E_{c.m.}}\sum_{J=0}^{J_{max}}(2J+1)
T(E_{c.m.},J)P_{CN}(E_{c.m.},J)W_{sur}(E_{c.m.},J).
\end{equation}
Here, $T(E_{c.m.},J)$ is the transmission probability of the two
colliding nuclei overcoming the Coulomb potential barrier in the
entrance channel to form the DNS. In the same manner as in the
nucleon collectivization model $\cite{Za01}$, the transmission
probability $T$ is calculated by using the empirical coupled channel
model, which can reproduce very well available experimental capture
cross sections $\cite{Za01,Fe06}$. The $P_{CN}$ is the probability
that the system will evolve from a touching configuration into the
compound nucleus in competition with quasi-fission of the DNS and
fission of the heavy fragment. The last term is the survival
probability of the formed compound nucleus, which can be estimated
with the statistical evaporation model by considering the
competition between neutron evaporation and fission $\cite{Fe06}$.
We take the maximal angular momentum as $J_{max}=30$ since the
fission barrier of the heavy nucleus disappears at high spin
$\cite{Re00}$.

In order to describe the fusion dynamics as a diffusion process in
mass asymmetry, the analytical solution of the Fokker-Planck
equation $\cite{Ad97}$ and the numerical solution of the master
equations $\cite{Li03,Di01}$ have been used, which were also used to
treat deep inelastic heavy-ion collisions $\cite{No75}$. Here, the
fusion probability is obtained by solving a set of master equations
numerically in the potential energy surface of the DNS. The time
evolution of the distribution function $P(A_{1},E_{1},t)$ for
fragment 1 with mass number $A_{1}$ and excitation energy $E_{1}$ is
described by the following master equations $\cite{Fe07,Li03}$,
\begin{eqnarray}
\frac{d P(A_{1},E_{1},t)}{dt}=\sum_{A_{1}^{\prime
}}W_{A_{1},A_{1}^{\prime}}(t)\left[
d_{A_{1}}P(A_{1}^{\prime},E_{1}^{\prime},t)-d_{A_{1}^{\prime
}}P(A_{1},E_{1},t)\right]-
\nonumber \\
\left[\Lambda^{qf}(\Theta(t))+\Lambda^{fis}(\Theta(t))
\right]P(A_{1},E_{1},t).
\end{eqnarray}
Here $W_{A_{1},A_{1}^{\prime}}$ is the mean transition probability
from the channel $(A_{1},E_{1})$ to
$(A_{1}^{\prime},E_{1}^{\prime})$, and $d_{A_{1}}$ denotes the
microscopic dimension corresponding to the macroscopic state
$(A_{1},E_{1})$. The sum is taken over all possible mass numbers
that fragment $A_{1}^{\prime}$ may take (from $0$ to
$A=A_{1}+A_{2}$), but only one nucleon transfer is considered in the
model with $A_{1}^{\prime }=A_{1}\pm 1$. The excitation energy
$E_{1}$ is the local excitation energy $\varepsilon^{\ast}_{1}$ with
respect to fragment $A_{1}$, which is determined by the dissipation
energy from the relative motion and the potential energy of the
corresponding DNS and will be shown later in Eqs.(8) and (9). The
dissipation energy is described by the parametrization method of the
classical deflection function $\cite{Wo78,Li81}$. The motion of
nucleons in the interacting potential is governed by the
single-particle Hamiltonian $\cite{Fe06,Li03}$:
\begin{equation}
H(t)=H_{0}(t)+V(t)
\end{equation}
with
\begin{eqnarray}
H_{0}(t)&=&\sum_{K}\sum_{\nu_{K}}\varepsilon_{\nu_{K}}(t)a_{\nu_{K}}^{\dag}(t)a_{\nu_{K}}(t),
\nonumber \\
V(t)&=&\sum_{K,K^{\prime}}\sum_{\alpha_{K},\beta_{K^{\prime}}}u_{\alpha_{K},\beta_{K^{\prime}}}(t)
a_{\alpha_{K}}^{\dag}(t)a_{\beta_{K^{\prime}}}(t)=\sum_{K,K^{\prime}}V_{K,K^{\prime}}(t).
\end{eqnarray}
Here the indices $K, K^{\prime}$ $(K,K^{\prime}=1,2)$ denote the
fragments $1$ and $2$. The quantities $\varepsilon_{\nu_{K}}$ and
$u_{\alpha_{K},\beta_{K^{\prime}}}$ represent the single particle
energies and the interaction matrix elements, respectively. The
single particle states are defined with respect to the centers of
the interacting nuclei and are assumed to be orthogonalized in the
overlap region. So the annihilation and creation operators are
dependent on time. The single particle matrix elements are
parameterized by
\begin{equation}
u_{\alpha_{K},\beta_{K^{\prime}}}(t)=U_{K,K^{\prime}}(t)\left\{\exp\left[-\frac{1}{2}
\left(\frac{\varepsilon_{\alpha_{K}}(t)-\varepsilon_{\beta_{K^{\prime}}}(t)}
{\Delta_{K,K^{\prime}}(t)}\right)^{2}\right]-
\delta_{\alpha_{K},\beta_{K^{\prime}}}\right\},
\end{equation}
which contain some parameters $U_{K,K^{\prime}}(t)$ and
$\Delta_{K,K^{\prime}}(t)$. The detailed calculation of these
parameters and the mean transition probabilities were described in
Refs. $\cite{Fe06,Li03}$.

The evolution of the DNS along the variable R leads to the
quasi-fission of the DNS. The quasi-fission rate $\Lambda^{qf}$ can
be estimated with the one-dimensional Kramers formula
$\cite{Ad03,Gr83}$:
\begin{equation}
\Lambda^{qf}(\Theta(t))=\frac{\omega}{2\pi\omega^{B_{qf}}}\left(\sqrt{\left(\frac{\Gamma}
{2\hbar}\right)^{2}+(\omega^{B_{qf}})^{2}}-\frac{\Gamma}
{2\hbar}\right)\exp\left(-\frac{B_{qf}(A_{1},A_{2})}{\Theta(t)}\right).
\end{equation}
Here the quasi-fission barrier is counted from the depth of the
pocket of the interaction potential. The local temperature is given
by the Fermi-gas expression $\Theta=\sqrt{\varepsilon^{\star}/a}$
corresponding to the local excitation energy $\varepsilon^{\star}$
and level density parameter $a=A/12$ $MeV^{-1}$. In Eq.(6) the
frequency $\omega^{B_{qf}}$ is the frequency of the inverted
harmonic oscillator approximating the interaction potential of two
nuclei in R around the top of the quasi-fission barrier, and
$\omega$ is the frequency of the harmonic oscillator approximating
the potential in R around the bottom of the pocket. The quantity
$\Gamma$, which denotes the double average width of the contributing
single-particle states, determines the friction coefficients:
$\gamma_{ii^{\prime}}=\frac{\Gamma}{\hbar}\mu_{ii^{\prime}}$, with
$\mu_{ii^{\prime}}$ being the inertia tensor. Here we use constant
values $\Gamma=2.8$ MeV, $\hbar\omega^{B_{qf}}=2.0$ MeV and
$\hbar\omega=3.0$ MeV for the following reactions. The Kramers
formula is derived with the quasi-stationary condition of the
temperature $\Theta(t)<B_{qf}(A_{1},A_{2})$. However, the numerical
calculation in Ref. $\cite{Gr83}$ indicated that Eq.(6) is also
useful for the condition of $\Theta(t)>B_{qf}(A_{1},A_{2})$. In the
reactions of synthesizing SHN, there is the possibility of the
fission of the heavy fragment in the DNS. Because the fissility
increases with the charge number of the nucleus, the fission of the
heavy fragment can affect the quasi-fission and fusion when the DNS
evolves towards larger mass asymmetry. The fission rate
$\Lambda^{fis}$ can also be treated with the one-dimensional Kramers
formula $\cite{Ad03}$
\begin{equation}
\Lambda^{fis}(\Theta(t))=\frac{\omega_{g.s.}}{2\pi\omega_{f}}\left(\sqrt{\left(\frac{\Gamma_{0}}
{2\hbar}\right)^{2}+\omega_{f}^{2}}-\frac{\Gamma_{0}}
{2\hbar}\right)\exp\left(-\frac{B_{f}(A_{1},A_{2})}{\Theta(t)}\right),
\end{equation}
where the $\omega_{g.s.}$ and $\omega_{f}$ are the frequencies of
the oscillators approximating the fission-path potential at the
ground state and on the top of the fission barrier for nucleus
$A_{1}$ or $A_{2}$ (larger fragment), respectively. Here, we take
$\hbar\omega_{g.s.}=\hbar\omega_{f}=1.0$ MeV, $\Gamma_{0}=2$ MeV.
The fission barrier is calculated as the sum of a macroscopic part
and the shell correction energy used in Refs. $\cite{Fe06,Ad00}$.
The fission of the heavy fragment does not favor the diffusion of
the system to a light fragment distribution. Therefore, it leads to
a slight decrease of the fusion probability.

In the relaxation process of the relative motion, the DNS will be
excited by the dissipation of the relative kinetic energy. The
excited system opens a valence space $\Delta\varepsilon_{K}$ in
fragment $K (K=1,2)$, which has a symmetrical distribution around
the Fermi surface. Only the particles in the states within this
valence space are actively involved in excitation and transfer. The
averages on these quantities are performed in the valence space:
\begin{equation}
\Delta\varepsilon_{K}=\sqrt{\frac{4\varepsilon^{\ast}_{K}}{g_{K}}},
\varepsilon^{\ast}_{K}=\varepsilon^{\ast}\frac{A_{K}}{A},
g_{K}=\frac{A_{K}}{12},
\end{equation}
where the $\varepsilon^{\ast}$ is the local excitation energy of the
DNS, which provides the excitation energy for the mean transition
probability. There are $N_{K}=g_{K}\Delta\varepsilon_{K}$ valence
states and $m_{K}=N_{K}/2$ valence nucleons in the valence space
$\Delta\varepsilon_{K}$, which gives the dimension
$d(m_{1},m_{2})=\left(\begin{array}{c}
N_{1} \\
m_{1}
\end{array}\right)
\left(\begin{array}{c}
N_{2} \\
m_{2}
\end{array}\right)$. The local
excitation energy is defined as
\begin{equation}
\varepsilon^{\ast}=E_{x}-\left(U(A_{1},A_{2})-U(A_{P},A_{T})\right).
\end{equation}
Here the $U(A_{1},A_{2})$ and $U(A_{P},A_{T})$ are the driving
potentials of fragments $A_{1}$, $A_{2}$ and fragments $A_{P}$,
$A_{T}$ (at the entrance point of the DNS), respectively. The
detailed calculation of the driving potentials can be seen in Ref.
$\cite{Fe07}$. The excitation energy $E_{x}$ of the composite system
is converted from the relative kinetic energy loss, which is related
to the Coulomb barrier $B$ $\cite{Fe07a}$ and determined for each
initial relative angular momentum $J$ by the parametrization method
of the classical deflection function $\cite{Wo78,Li81}$. So $E_{x}$
is coupled with the relative angular momentum.

After reaching the reaction time in the evolution of
$P(A_{1},E_{1},t)$, all those components on the left side of the
B.G. (Businaro-Gallone) point contribute to the formation of the
compound nucleus. The hindrance in the diffusion process by nucleon
transfer to form the compound nucleus is the inner fusion barrier
$B_{fus}$, which is defined as the difference of the driving
potential at the B.G. point and at the entrance position. Nucleon
transfers to more symmetric fragments undergo quasi-fission. The
formation probability of the compound nucleus at the Coulomb barrier
$B$ (here a barrier distribution $f(B)$ is considered) and angular
momentum $J$ is given by
\begin{equation}
P_{CN}(E_{c.m.},J,B)=\sum_{A_{1}=1}^{A_{BG}}P(A_{1},E_{1},\tau
_{int}(E_{c.m.},J,B)).
\end{equation}
Here the interaction time $\tau _{int}(E_{c.m.},J,B)$ is obtained
using the deflection function method $\cite{Li83}$, which means the
time duration for nucleon transfer from the capture stage to the
formation of the complete fused system with the order of 10$^{-20}$
s. We obtain the fusion probability as
\begin{equation}
P_{CN}(E_{c.m.},J)=\int f(B)P_{CN}(E_{c.m.},J,B)dB,
\end{equation}
where the barrier distribution function is taken in asymmetric
Gaussian form $\cite{Za01,Fe06}$. So the fusion cross section is
written as
\begin{equation}
\sigma_{fus}(E_{c.m.})=\frac{\pi \hbar^{2}}{2\mu
E_{c.m.}}\sum_{J=0}^{\infty}(2J+1) T(E_{c.m.},J)P_{CN}(E_{c.m.},J).
\end{equation}

The survival probability of the excited compound nucleus cooled by
the neutron evaporation in competition with fission is expressed as
follows:
\begin{equation}
W_{sur}(E_{CN}^{\ast},x,J)=P(E_{CN}^{\ast},x,J)\prod\limits_{i=1}^{x}\left(
\frac{\Gamma _{n}(E_{i}^{\ast},J)}{\Gamma
_{n}(E_{i}^{\ast},J)+\Gamma _{f}(E_{i}^{\ast},J)}\right) _{i},
\end{equation}
where the $E_{CN}^{\ast}, J$ are the excitation energy and the spin
of the compound nucleus, respectively. The $E_{i}^{\ast}$ is the
excitation energy before evaporating the $i$th neutron, which has
the relation
\begin{equation}
E_{i+1}^{\ast}=E_{i}^{\ast}-B_{i}^{n}-2T_{i},
\end{equation}
with the initial condition $E_{1}^{\ast}=E_{CN}^{\ast}$. The energy
$B_{i}^{n}$ is the separation energy of the $i$th neutron. The
nuclear temperature $T_{i}$ is given by
$E_{i}^{\ast}=aT_{i}^{2}-T_{i}$ with the level density parameter
$a$. $P(E_{CN}^{\ast},x,J)$ is the realization probability of
emitting $x$ neutrons. The widths of neutron evaporation and fission
are calculated using the statistical model. The details can be found
in Ref. $\cite{Fe06}$. The level density is expressed by the
back-shifted Bethe formula $\cite{Be36}$ with the spin cut-off model
as
\begin{equation}
\rho(E^{\ast},J)=K_{rot}K_{vib}\frac{2J+1}{24\sqrt{2}\sigma^{3}}a^{-1/4}(E^{\ast}-\Delta)^{-5/4}
\exp[2\sqrt{a(E^{\ast}-\Delta)}]\exp[-\frac{(J+1/2)^{2}}{2\sigma^{2}}],
\end{equation}
where the $K_{rot}$ and $K_{vib}$ are the coefficients of the
rotational and vibrational enhancements. The pairing energy is given
by
\begin{equation}
\Delta=\chi\frac{12}{\sqrt{A}}
\end{equation}
in MeV($\chi$=-1, 0 and 1 for odd-odd, odd-even and even-even
nuclei, respectively). The spin cut-off parameter is calculated by
the formula:
\begin{equation}
\sigma^{2}=T\zeta_{r.b}/\hbar^{2},
\end{equation}
where the rigid-body moment of inertia has the relation
$\zeta_{r.b}=0.4MR^{2}$ with the mass $M$ and the radius $R$ of the
nucleus. The level density parameter is related to the shell
correction energy $E_{sh}(Z,N)$ and the excitation energy $E^{\ast}$
of the nucleus as
\begin{equation}
a(E^{\ast},Z,N)=\tilde{a}(A)[1+E_{sh}(Z,N)f(E^{\ast}-\Delta)/(E^{\ast}-\Delta)].
\end{equation}
Here, $\tilde{a}(A)=\alpha A+\beta A^{2/3}b_{s}$ is the asymptotic
Fermi-gas value of the level density parameter at high excitation
energy. The shell damping factor is given by
\begin{equation}
f(E^{\ast})=1-\exp(-\gamma E^{\ast})
\end{equation}
with $\gamma=\tilde{a}/(\epsilon A^{4/3})$. All the used parameters
are listed in Table 1. In Fig.1 we give the level density parameters
of different nuclides at the ground state calculated by using
Eq.(18) and compared them with two empirical formulas $a(A)=A/8$,
and $A/12$. It can be seen that the strong shell effects appear in
the level density.

With this procedure introduced above, we calculated the angular
momentum dependence of the capture, fusion and survival
probabilities as shown in Fig.2 for the reaction
$^{48}$Ca+$^{208}$Pb at incident energies 172.36 MeV and 192.36 MeV,
respectively. The values of the three stages decrease obviously with
increasing the relative angular momentum. So in the following
estimation of the production cross sections, we cut off the maximal
angular momentum at $J_{max}=30$, which is taken as the same value
that used in the cold fusion reactions $\cite{Fe07}$.

\section{Results and discussions}
\subsection{Fusion-fission reactions and quasi-fission mass yields}

As a test of the parameters for the estimation of the transmission
of two colliding nuclei and the de-excitation of the thermal
compound nucleus, we analyzed the fusion-fission reactions for the
selected systems shown in Fig.3 assuming $P_{CN}=1$. The capture and
evaporation residue cross sections are compared with the available
experimental data $\cite{Pr08,Da04,Ni04,Ga08}$. For these systems
the quasi-fission does not dominate in the sub-barrier region, which
also means that $P_{CN}\sim1$. The evaporation residues are mainly
determined through the capture of the light projectile by the target
nucleus and the survival probabilities of the formed compound
nucleus. The experimental data can be reproduced rather well within
the error bars. Some discrepancies may come from the quasi-fission
in the above barrier region and from the input quantities, such as
the neutron separation energy, shell correction and mass. The
rotational and the vibrational enhancement in the level density can
also affect the survival probabilities of the excited compound
nucleus $\cite{Ju98}$. Here we take unity for both coefficients as
shown in Table 1 because the height of the fission barrier is also
sensitive to the survival of the compound nucleus by fitting the
experimental evaporation residue excitation functions in the
fusion-fission reactions.

Since the electrostatic energy of the composite systems formed by
two heavy colliding nuclei is very large, so although the two nuclei
may be captured by the nuclear potential, they almost always
separate after mass transfer from the heavier nucleus to the lighter
one rather fusing. This process is called quasi-fission
$\cite{Ba85,To85}$, which is the main feature in the massive fusion
reactions and can inhibit fusion by several degree of freedom.
Recently, experiment has performed nice works by measuring the
quasi-fission and fusion-fission mass yields $\cite{It04}$. In the
DNS model, the quasi-fission mass yields are expressed as
$\cite{Ad03}$
\begin{equation}
Y_{q-f}(A_{1})=\sum_{J=0}^{J_{max}}\int_{0}^{\tau_{int}}
P(A_{1},E_{1},t)\Lambda^{qf}(\Theta(t))dt.
\end{equation}
In Fig.4 we show a comparison of the calculated quasi-fission mass
yields and the experimental data for the two $^{48}$Ca induced
reaction systems. The trends of the distribution can be reproduced
by the DNS model. At the domain of the medium-mass fragments
A$_{1}$=A$_{CN}$/2-30$\sim$A$_{CN}$/2+30, The experimental data are
higher than the calculated values, which may be come from the
contribution of the fusion-fission fragments.

\subsection{Evaporation residue cross sections}

The evaporation residues observed in laboratories by the consecutive
$\alpha$ decay are mainly produced by the complete fusion reactions,
in which the fusion dynamics and the structure properties of the
compound nucleus affect their production. Within the framework of
the DNS model, we calculated the evaporation residue cross sections
producing SHN Z=110, 112, 113, 115 with $^{232}$Th, $^{238}$U,
$^{237}$Np and $^{243}$Am targets in the $^{48}$Ca induced reactions
as shown in Fig.5, and compared them with the Dubna data
$\cite{Og04a,Og04b,Og07a}$ as well as with the recent GSI data
$\cite{Ho07}$ for $^{238}$U targets in the 3n channel. Compared with
the Dubna data for the system $^{48}$Ca+$^{238}$U, the GSI results
show that the formation cross sections in the 3n channel have a
slight decrease at the same excitation energy, which is in a good
agreement with our calculated results. The calculations were carried
out before getting the experimental data $\cite{Og07a}$ for the
reaction $^{48}$Ca+$^{237}$Np, and a good agreement with the data is
also found $\cite{Fe07b}$. The excitation energy of the compound
nucleus is obtained by $E^{\ast}_{CN}=E_{c.m.}+Q$, where the
$E_{c.m.}$ is the incident energy in the center-of-mass system. The
$Q$ value is given by $Q=\Delta M_{P}+\Delta M_{T}-\Delta M_{C}$,
and the corresponding mass excesses $\Delta M_{i}$ ($i=P, T, C$) are
taken the data from Ref. $\cite{Mo95}$ for the projectile, target
and compound nucleus denoted with the symbols $P$, $T$ and $C$,
respectively. Usually, the neutron-rich projectile-target
combinations are in favor of synthesizing SHN experimentally, which
can enhance the survival probability $W_{sur}$ in Eq.(1) of the
formed compound nucleus because of the smaller neutron separation
energy. Differently to the cold fusion reactions $\cite{Fe07}$, the
maximal production cross sections from Ds to 115 especially in the
2n-5n channels are not changed much although the heavier SHNs are
synthesized. Within the error bars the experimental data can be
reproduced rather well. With the same procedure, we analyzed the
evaporation residue excitation functions with targets $^{242,244}$Pu
and $^{245,248}$Cm that are used to synthesize the superheavy
elements Z=114 and 116 in Dubna $\cite{Og04b,Og04c}$ (Fig.6). Our
calculations show that the target $^{244}$Pu has a larger production
cross section than $^{242}$Pu because of the larger survival
probability. In Fig.7 we also calculated the evaporation residue
excitation functions to synthesize superheavy elements Z=117-120
using the actinide isotopes with longer half-lives $^{247}$Bk,
$^{249}$Cf, $^{254}$Es and $^{257}$Fm. The 3n evaporation channel
with an excitation energy of the formed compound nucleus around 30
MeV is favorable to produce SHN with Z$\geq$117 by using the
actinide targets. Within the error bars, the positions of the
maximal production cross sections are in good agreement with the
available experimental results. Similar calculation of the
evaporation residue excitation functions was also reported in Ref.
$\cite{Za04}$. The spectrum form of evaporating neutrons is mainly
determined by the survival probability, in which the neutron
separation energy and the shell correction play a very important
role in the determination of the value. We considered the angular
momentum influence in the calculation of the level density, but did
not include it in the estimation of the fission barrier of the
thermal compound nucleus. As pointed out in section 1, the fission
barrier of SHN decreases rapidly with increasing excitation energy
of the compound nucleus, where the rotation of the system affects
the height of the barrier and also influences other crucial
quantities such as the level density etc.

In Fig.8 we show a comparison of the calculated maximal production
cross sections of superheavy elements Z=102-120 in the cold fusion
reactions by evaporating one neutron, in the $^{48}$Ca induced
reactions with actinide targets by evaporating three neutrons, and
the experimental data $\cite{Ho00,Og07,Mu99,Gr03}$. The production
cross sections decrease rapidly with increasing the charge number of
the synthesized compound nucleus in the cold fusion reactions, such
as from 0.2 $\mu b$ for the reaction $^{48}$Ca+$^{208}$Pb to 1 pb
for $^{70}$Zn+$^{208}$Pb, and even below 0.1 pb for synthesizing
Z$\geq$113 $\cite{Fe07}$. It seems to be difficult to synthesize
superheavy elements Z$\geq$113 in the cold fusion reactions at the
present facilities. The calculated results show that the $^{48}$Ca
induced reactions have smaller production cross sections with
$^{232}$Th target, but are in favor of synthesizing heavier SHN
(Z$\geq$113) because of the larger cross sections. The experimental
data also give such trends. In the DNS concept, the inner fusion
barrier increases with reducing mass asymmetry in the cold fusion
reactions, which leads to a decrease of the formation probability of
the compound nucleus. However, the $^{48}$Ca induced reactions have
not such increase of the inner fusion barrier for synthesizing
heavier SHN. Because of the larger transmission and the higher
fusion probability, we obtain larger production cross sections for
synthesizing SHN (Z$\geq$113) in the $^{48}$Ca induced reactions
although these reactions have the smaller survival probability than
those in the cold fusion reactions. It is still a good way to
synthesize heavier SHN by using the $^{48}$Ca induced reactions. Of
course, further experimental data are anticipated to be obtained in
the future. However, the actinide targets are difficulty to be
handled in experiments synthesizing heavier SHN.

\subsection{Isotopic dependence of the production cross sections}

Recent experimental data show that the production cross sections of
the SHN depend on the isotopic combination of the target and
projectile in the $^{48}$Ca induced fusion reactions. For example,
the maximal cross section in the 3n channel is $3.7\pm^{3.6}_{1.8}$
pb for the reaction $^{48}$Ca+$^{245}$Cm at the excitation energy
37.9 MeV; however, it is $1.2$ pb for the reaction
$^{48}$Ca+$^{248}$Cm although the later is a neutron-rich target
$\cite{Og06,Og04b}$. The isotopic trends of the production cross
sections were also observed and investigated in cold fusion
reactions $\cite{Ho95,Fe07}$. Further investigations on the isotopic
trends in the $^{48}$Ca induced reactions are very necessary for
predicting the optimal combinations, excitation energies (incident
energies) and evaporation channels in the synthesis of SHN. In Fig.9
we show the calculated isotopic trends in producing superheavy
elements Z=110, 112 with the isotopic actinides Th and U in the 3n
channels, and compare them with the available experimental data
performed in Dubna $\cite{Og04b}$ (squares with error bars) and at
GSI $\cite{Ho07}$ (circles with error bars). The results show that
the targets $^{230}$Th in the 4n channel and $^{235,238}$U in the 3n
channel have the largest cross sections. The isotopic trends in
synthesizing Z=113-116 with the actinide targets Np, Pu, Am and Cm
are also calculated systematically, and compared with the existing
data measured in Dubna $\cite{Og04a,Og04b,Og04c}$ and the results of
Adamian et al. $\cite{Ad04}$ for the Pu isotopes as shown in Fig.10
and Fig.11. The isotopes $^{237}$Np, $^{241}$Pu, $^{242,243}$Am and
$^{245,247}$Cm in the 3n channels, and $^{244}$Pu in the 4n channel
as well as the isotope $^{250}$Cm are suitable for synthesizing SHN.
Except for the $^{244}$Pu, our calculated cross sections are smaller
than the ones of the Adamian et al. In the DNS model, the isotopic
dependence of the production cross sections is mainly determined by
both the fusion and survival probabilities. Of course, the
transmission probability of two colliding nuclei can also be
affected since the isotopes have initial quadrupole deformations.
With the same procedure, we analyzed the dependence of the
production cross sections on the isotopes Bk and Cf in the 3n
channels for synthesizing the superheavy elements Z=117, 118 and
compared them with the available experimental data $\cite{Og06}$
shown in Fig.12. The results show that the targets $^{248,249}$Bk
and $^{251,252}$Cf are favorable for synthesizing the superheavy
elements Z=117 and 118. The corresponding excitation energies are
also given in the figures.

In Fig.13 we show the dependence of the inner fusion barrier, the
fission barrier of the compound nucleus, and the neutron separation
energies of evaporating 3n and 4n on the mass numbers of the
isotopic targets Cm in the $^{48}$Ca induced reactions. It is
obvious that the combinations with the isotopes $^{245,247}$Cm have
smaller inner fusion barriers, higher fission barriers and smaller
3n separation energies, which result in larger production cross
sections producing the superheavy element Z=116. Although the lower
fission barrier for the isotope $^{250}$Cm, it gives the smaller
inner fusion barrier and neutron separation energies, which also
leads to the larger cross sections in the 3n and 4n channels as
shown in Fig.11. The shell correction and the neutron separation
energies are taken from Ref. $\cite{Mo95}$. When the neutron number
of the target increases, the DNS gets more asymmetrical and the
fusion probability increases if the DNS does not consist of more
stable nuclei (such as magic nuclei) because of a smaller inner
fusion barrier. A smaller neutron separation energy and a larger
shell correction lead to a larger survival probability. The compound
nucleus with closed neutron shells has a larger shell correction
energy and a larger neutron separation energy. The neutron-rich
actinide target has larger fusion and survival probabilities due to
the larger asymmetric initial combinations and smaller neutron
separation energies. But such actinide isotopes are usually unstable
with smaller half-lives. With the establishment of the high
intensity radioactive-beam facilities, the neutron-rich SHN may be
synthesized experimentally, which approaches the island of
stability.

\subsection{$^{238}$U based reactions}

The uranium is the heaviest element existing in the nature. It has a
larger mass asymmetry constructed as a target in the fusion
reactions with the various neutron-rich light projectiles. The
isotope $^{238}$U is the neutron-richest nucleus in the U isotopes
and often chosen as the target for synthesizing SHN. In Fig.14 we
give evaporation residue excitation functions of the reactions
$^{40}$Ar, $^{50}$Ti, $^{54}$Cr, $^{64}$Ni+$^{238}$U in the 2n-5n
channels. The results show that the 4n channel in the reaction
$^{40}$Ar+$^{238}$U has the larger cross sections with 2.1 pb at an
excitation energy 42 MeV. This reaction is being used to synthesize
the superheavy nucleus Ds with HIRFL accelerator at Institute of
Modern Physics in Lanzhou. The reactions $^{50}$Ti, $^{54}$Cr,
$^{64}$Ni+$^{238}$U lead to the cross section smaller than 0.1 pb.
The isotopic trends based on the U isotopes are also investigated
using the DNS model as shown in Fig.15. Calculations show that the
isotopes $^{235}$U and $^{238}$U are favorable in producing SHN. The
cross sections are reduced with increasing the mass numbers of the
projectiles. Other reaction mechanisms to synthesize SHN have to be
investigated with theoretical models, such as the massive transfer
reactions, and the complete fusion reactions induced by weakly bound
nuclei. Work in these directions is in progress within the framework
of the DNS model.

\section{Conclusions}

Using the DNS model, we systematically investigated the production
of superheavy residues in fusion-evaporation reactions, in which the
nucleon transfer leading to the formation of the superheavy compound
nucleus is described with a set of microscopically derived master
equations that are solved numerically and include the quasi-fission
of the DNS and the fission of the heavy fragments. The fusion
dynamics and the evaporation residue excitation functions in the
$^{48}$Ca fusion reactions are systematically investigated. The
calculated results are in good agreement with the available
experimental data within the error bars. Isotopic trends in the
production of superheavy elements are analyzed. It is shown that the
isotopes $^{235,238}$U, $^{237}$Np, $^{241,244}$Pu, $^{242}$Am and
$^{245,247,250}$Cm, $^{248,249}$Bk and $^{251,252}$Cf in the 3n
channels, and $^{230}$Th, $^{244}$Pu, $^{248,250}$Cm in the 4n
channels are favorable for producing the superheavy elements Z=110,
112 and 113-118, respectively. The evaporation residue excitation
functions of the reactions $^{40}$Ar, $^{50}$Ti, $^{54}$Cr,
$^{64}$Ni+$^{238}$U in the 2n-5n channels and the isotopic trends
with $^{40}$Ar, $^{48}$Ca, $^{50}$Ti, $^{54}$Cr, $^{58}$Fe and
$^{64}$Ni bombarding U isotopes are also studied.

\section{Acknowledgement}

One of us (Z.-Q. Feng) is grateful to Prof. H. Feldmeier, Dr. G.G.
Adamian and Dr. N.V. Antonenko for fruitful discussions and help,
and also thanks the hospitality during his stay in GSI. This work
was supported by the National Natural Science Foundation of China
under Grant No. 10805061, the special foundation of the president
fellowship, the west doctoral project of Chinese Academy of
Sciences, and major state basic research development program under
Grant No. 2007CB815000.

\newpage

\begin{table}
\tabcolsep 0pt \caption{Parameters used in the calculation of the
level density.} \vspace*{-12pt}
\begin{center}
\def\temptablewidth{0.6\textwidth}
{\rule{\temptablewidth}{1pt}}
\begin{tabular*}{\temptablewidth}{@{\extracolsep{\fill}}cccccc}
$K_{rot}$ &$K_{vib}$ &$b_{s}$ \quad&$\alpha$ \quad&$\beta$ \quad&$\epsilon$ \\
\hline 1 &1 &1 \quad&0.114 \quad&0.098 \quad&0.4
\end{tabular*}
{\rule{\temptablewidth}{1pt}}
\end{center}
\end{table}

\begin{figure}
\begin{center}
{\includegraphics*[width=0.6\textwidth]{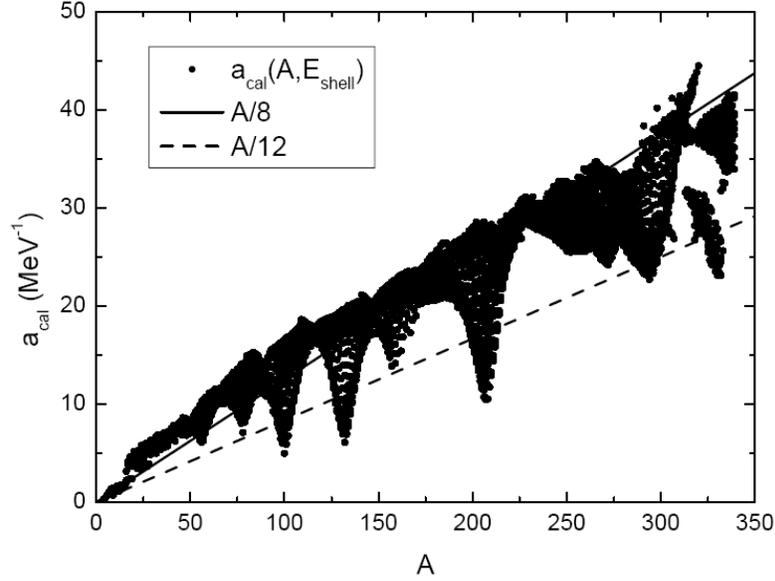}}
\end{center}
\caption{Calculated values of the level density parameters as a
function of the atomic mass.}
\end{figure}

\begin{figure}
\begin{center}
{\includegraphics*[width=0.9\textwidth]{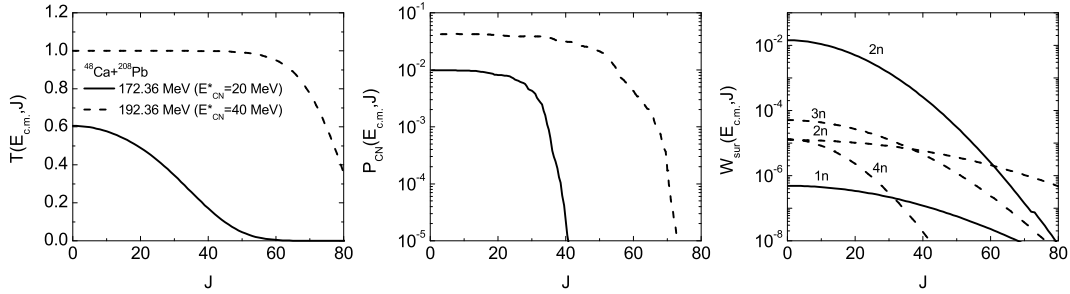}}
\end{center}
\caption{Calculated capture, fusion and survival probabilities as
functions of the relative angular momenta in the reaction
$^{48}$Ca+$^{208}$Pb at excitation energies of the compound nucleus
of 20 MeV and 40 MeV, respectively.}
\end{figure}

\begin{figure}
\begin{center}
{\includegraphics*[width=0.8\textwidth]{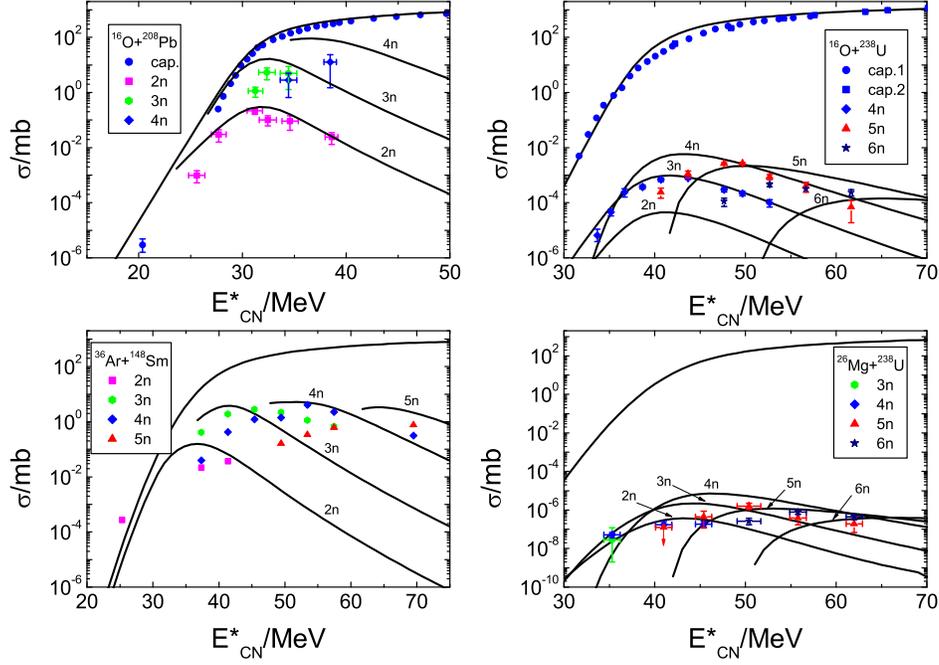}}
\end{center}
\caption{Comparison of the calculated fusion-fission excitation
functions and the available experimental data for the reactions
$^{16}$O+$^{208}$Pb, $^{16}$O+$^{238}$U, $^{36}$Ar+$^{148}$Sm and
$^{26}$Mg+$^{238}$U.}
\end{figure}

\begin{figure}
\begin{center}
{\includegraphics*[width=0.8\textwidth]{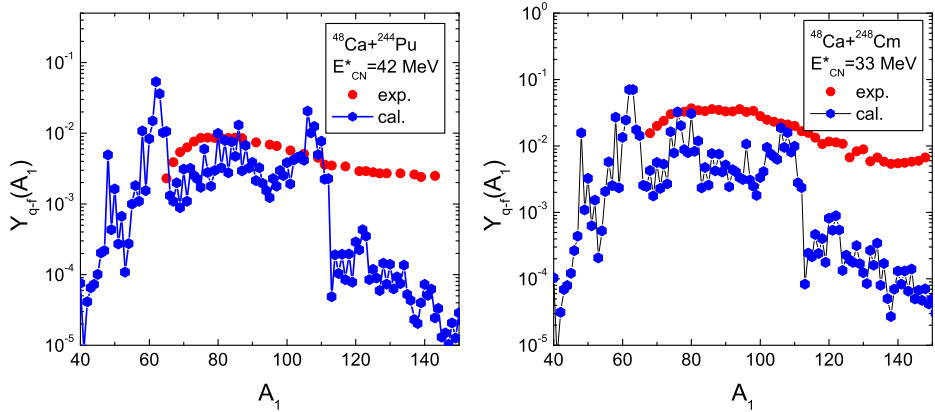}}
\end{center}
\caption{Calculated quasi-fission mass yields for the reactions
$^{48}$Ca+$^{244}$Pu and $^{48}$Ca+$^{248}$Cm at excitation energies
of the compound nuclei 42 MeV and 33 MeV, respectively, and compared
them with the available experimental data $\cite{It04}$.}
\end{figure}

\begin{figure}
\begin{center}
{\includegraphics*[width=0.8\textwidth]{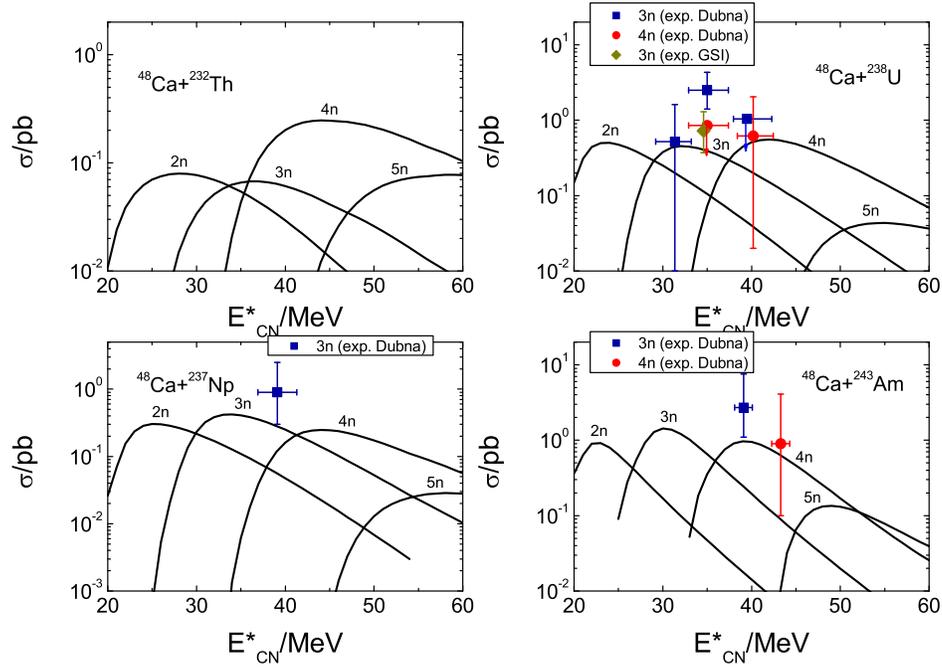}}
\end{center}
\caption{The calculated evaporation residue excitation functions
with $^{232}$Th, $^{238}$U, $^{237}$Np and $^{243}$Am targets in
$^{48}$Ca induced reactions, and compared with the available
experimental data $\cite{Og04a,Og04b,Og07a}$.}
\end{figure}

\begin{figure}
\begin{center}
{\includegraphics*[width=0.8\textwidth]{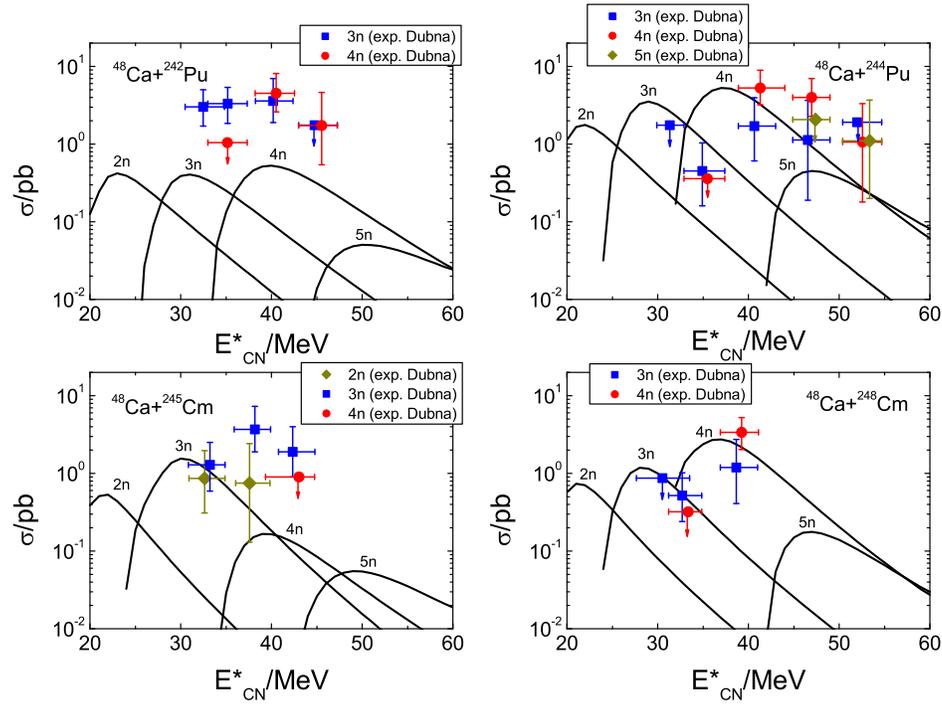}}
\end{center}
\caption{The same as in Fig.5, but for the targets $^{242,244}$Pu
and $^{245,248}$Cm to produce superheavy elements Z=114 and 116.}
\end{figure}

\begin{figure}
\begin{center}
{\includegraphics*[width=0.8\textwidth]{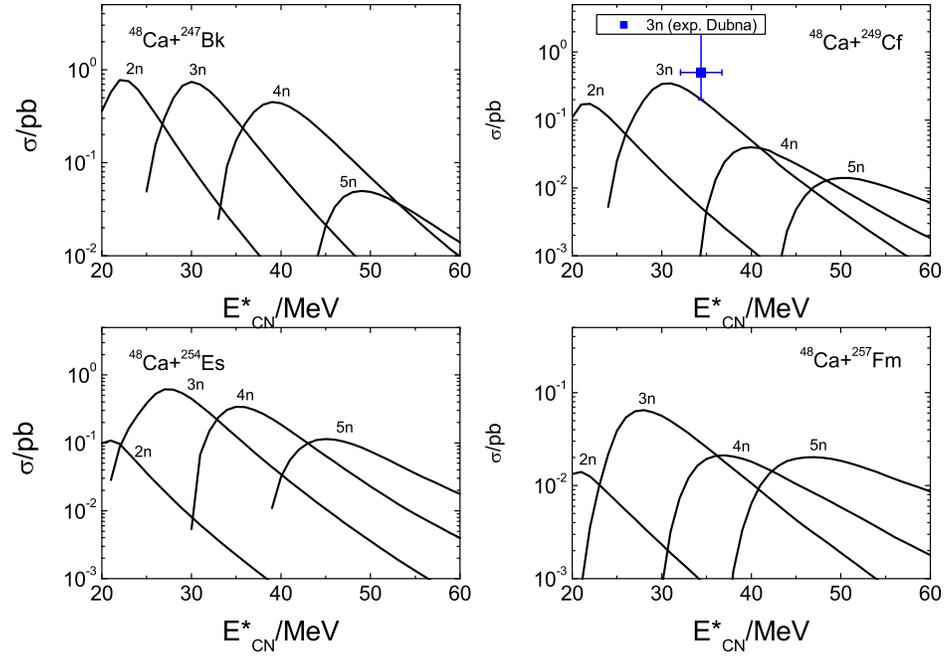}}
\end{center}
\caption{The same as in Fig.5, but for the targets $^{247}$Bk,
$^{249}$Cf, $^{254}$Es and $^{257}$Fm to synthesize superheavy
elements Z=117-120.}
\end{figure}

\begin{figure}
\begin{center}
{\includegraphics*[width=0.8\textwidth]{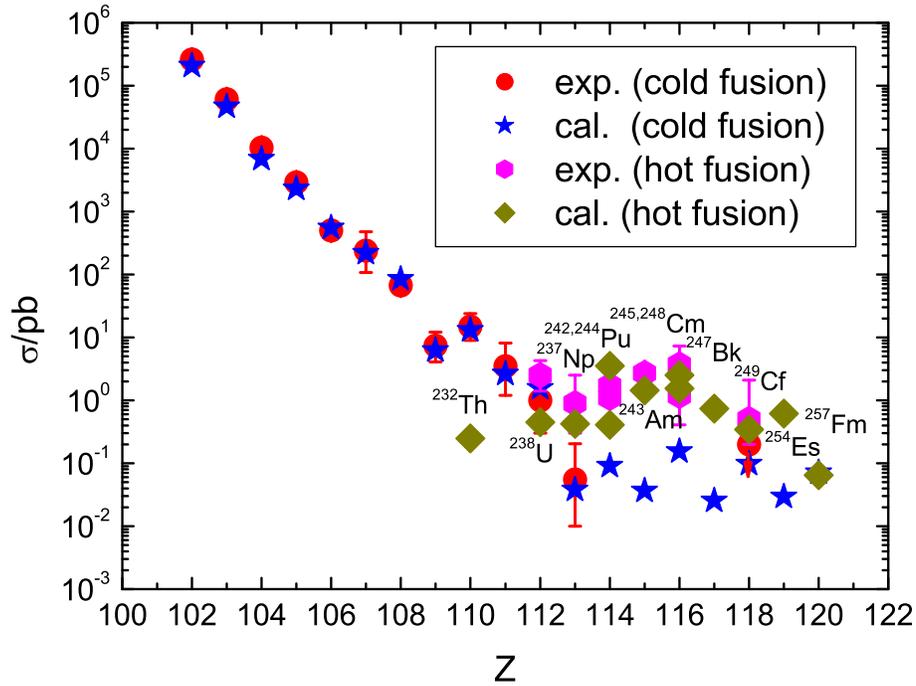}}
\end{center}
\caption{Maximal production cross sections of superheavy elements
Z=102-120 in cold fusion reactions based on $^{208}$Pb and
$^{209}$Bi targets with projectile nuclei $^{48}$Ca, $^{50}$Ti,
$^{54}$Cr, $^{58}$Fe, $^{64}$Ni, $^{70}$Zn, $^{76}$Ge, $^{82}$Se,
$^{86}$Kr and $^{88}$Sr, in $^{48}$Ca induced reactions with
actinide targets by evaporating 3 neutrons, in comparison with
available experimental data $\cite{Ho00,Og07,Mu99,Gr03}$.}
\end{figure}

\begin{figure}
\begin{center}
{\includegraphics*[width=0.8\textwidth]{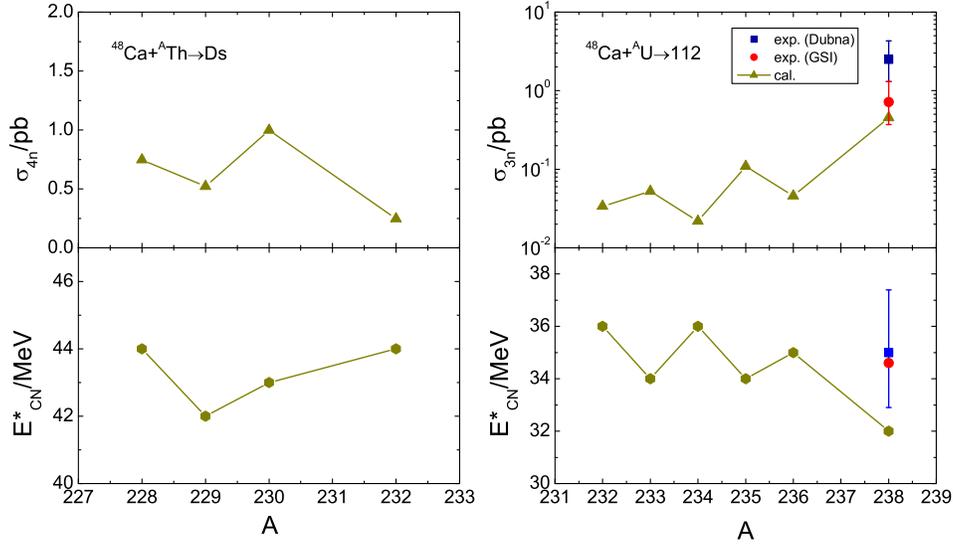}}
\end{center}
\caption{Isotopic dependence of the calculated maximal production
cross sections in the 3n evaporation channel and the corresponding
excitation energies in the synthesis of superheavy elements Z=110
and 112 for the reactions $^{48}$Ca+$^{A}$Th and $^{48}$Ca+$^{A}$U,
and compared with the experimental data $\cite{Og04b,Ho07}$.}
\end{figure}

\begin{figure}
\begin{center}
{\includegraphics*[width=0.8\textwidth]{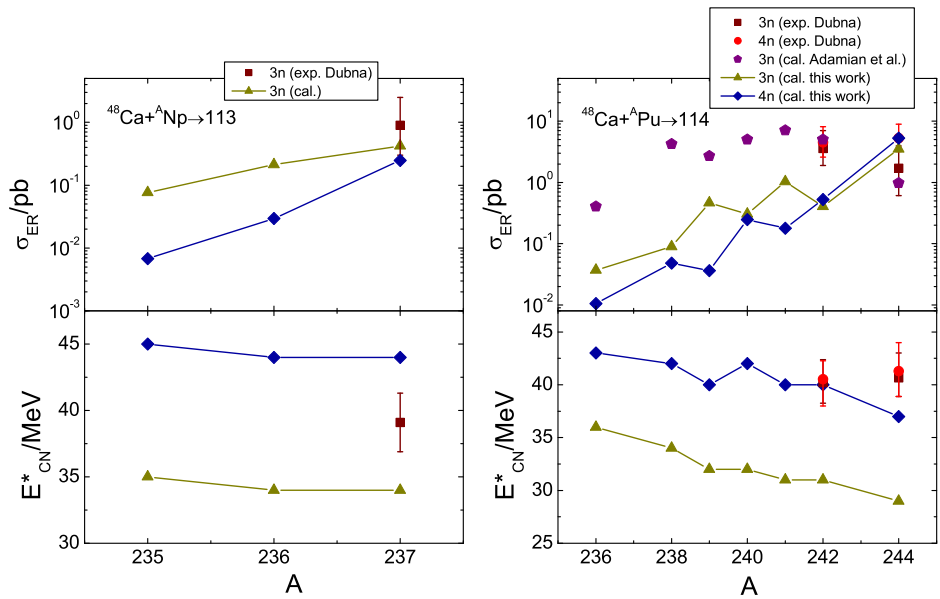}}
\end{center}
\caption{The same as in Fig.9, but for isotopic targets Np and Pu to
produce superheavy elements Z=113 and 114.}
\end{figure}

\begin{figure}
\begin{center}
{\includegraphics*[width=0.8\textwidth]{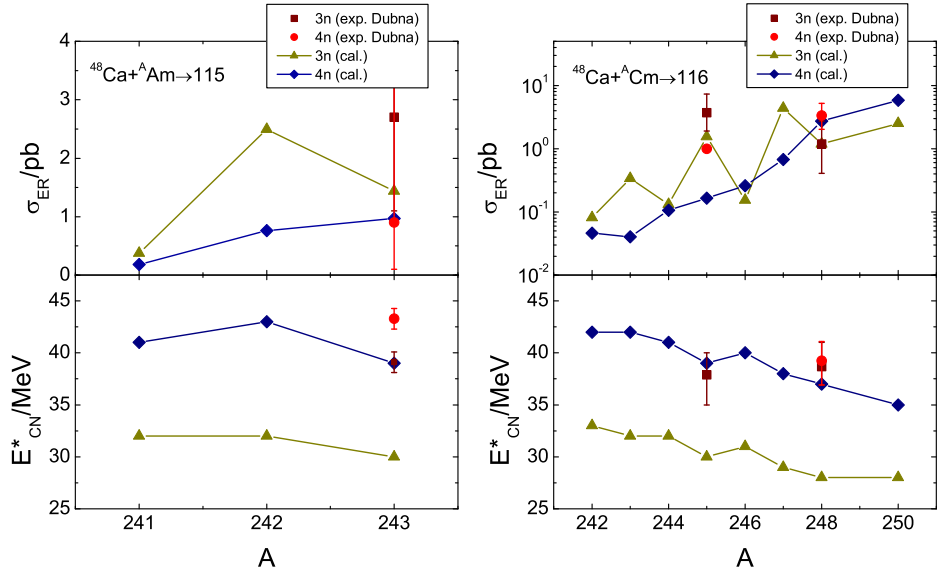}}
\end{center}
\caption{The same as in Fig.9, but for isotopic targets Am and Cm to
synthesize superheavy elements Z=115 and 116 in 3n and 4n channels.}
\end{figure}

\begin{figure}
\begin{center}
{\includegraphics*[width=0.8\textwidth]{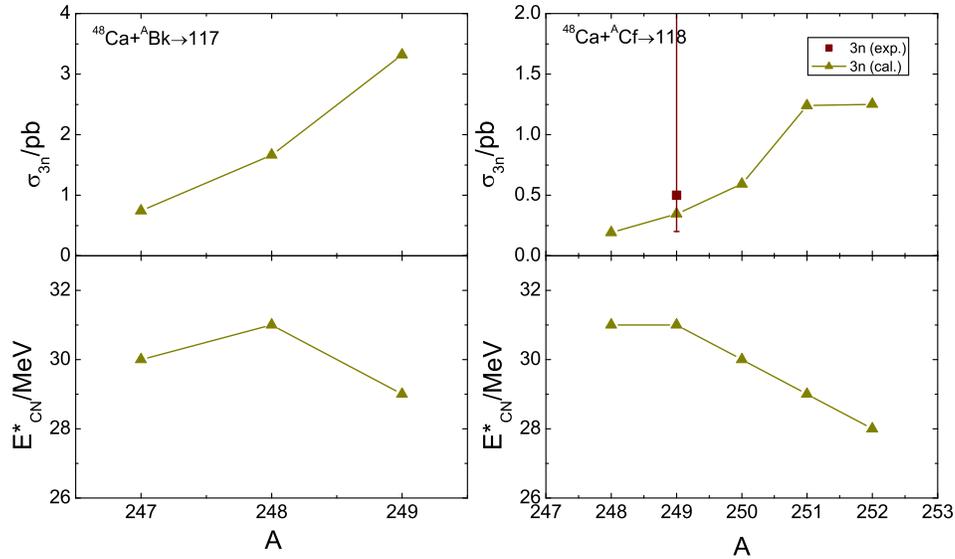}}
\end{center}
\caption{The same as in Fig.9, but for isotopes Bk and Cf in
$^{48}$Ca induced reactions.}
\end{figure}

\begin{figure}
\begin{center}
{\includegraphics*[width=0.8\textwidth]{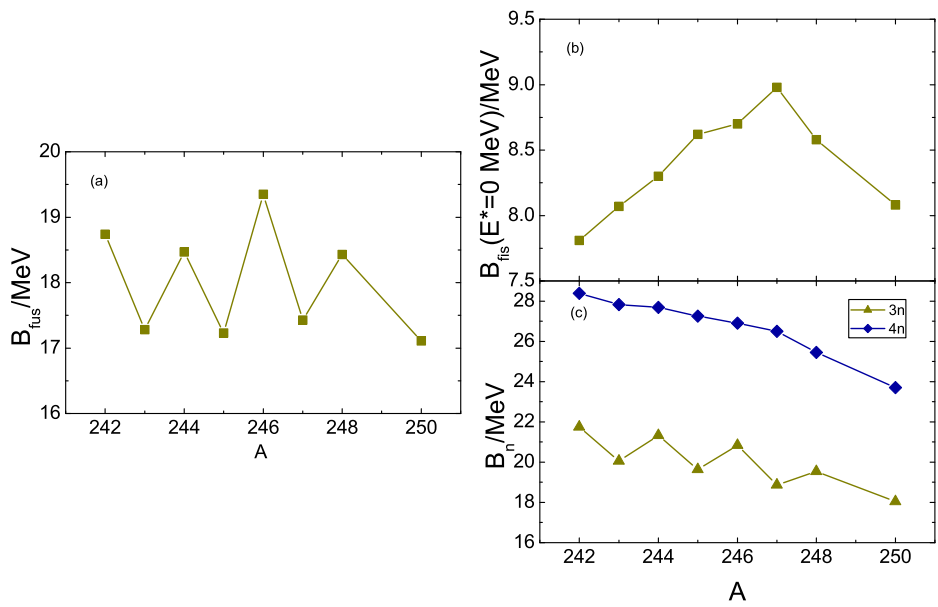}}
\end{center}
\caption{(a) the inner fusion barrier, (b) the fission barrier of
the compound nucleus and (c) the neutron separation energy as a
function of the mass numbers of the isotopic targets Cm in the
reactions $^{48}$Ca+$^{A}$Cm.}
\end{figure}

\begin{figure}
\begin{center}
{\includegraphics*[width=0.8\textwidth]{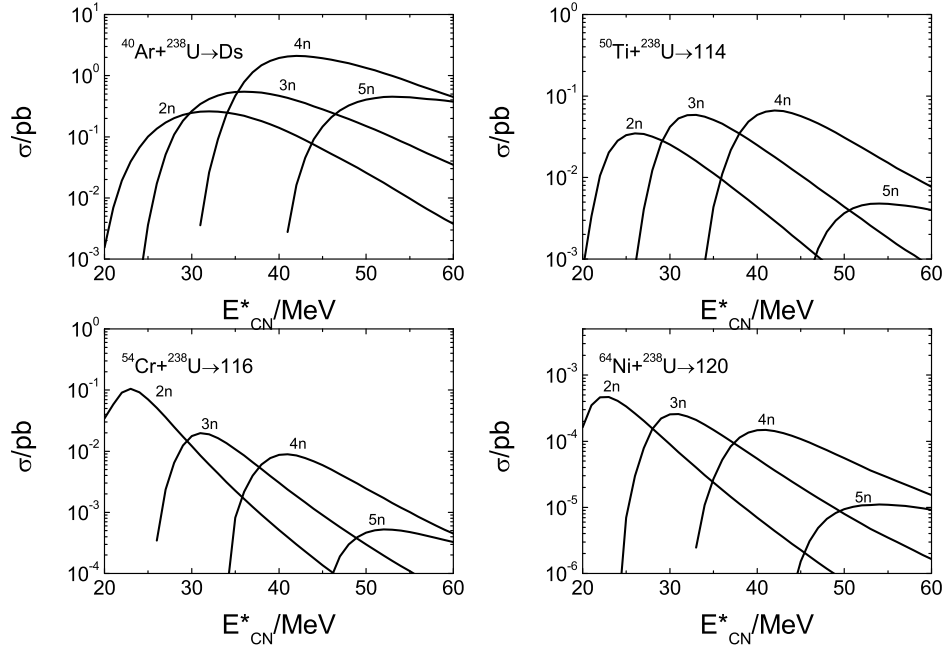}}
\end{center}
\caption{The evaporation residue excitation functions in the
reactions $^{40}$Ar, $^{50}$Ti, $^{54}$Cr, $^{64}$Ni+$^{238}$U.}
\end{figure}

\begin{figure}
\begin{center}
{\includegraphics*[width=0.8\textwidth]{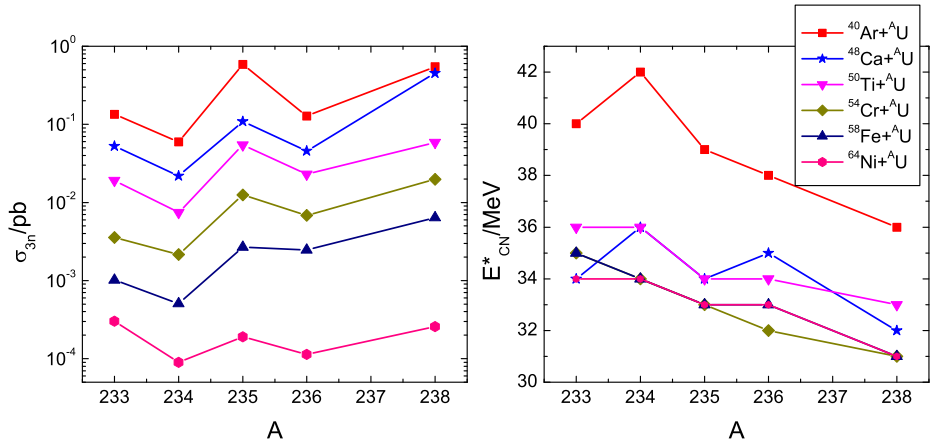}}
\end{center}
\caption{The production cross sections in the 3n channels as a
function of the mass number of the isotopic targets U with
projectiles $^{40}$Ar, $^{48}$Ca, $^{50}$Ti, $^{54}$Cr, $^{58}$Fe
and $^{64}$Ni.}
\end{figure}

\end{document}